\documentclass[aps,prl,twocolumn,groupedaddress,showpacs,superscriptaddress]{revtex4-1}
\usepackage{graphicx,amsmath}
\usepackage{amssymb}

\begin{document}
\title{Computational Supremacy of Quantum Eigensolver by Extension of Optimized Binary Configurations}
\author{Hayun Park}
\affiliation{Department of Liberal Studies, Kangwon National University, Samcheok, 25913, Republic of Korea}
\author{Hunpyo Lee}
\email{Email: hplee@kangwon.ac.kr}
\affiliation{Department of Liberal Studies, Kangwon National University, Samcheok, 25913, Republic of Korea}
\affiliation{Quantum Sub Inc., Samcheok, 25913, Republic of Korea}
\date{\today}

\begin{abstract}
We developed a quantum eigensolver (QE) which is based on an extension of optimized binary 
configurations measured by quantum annealing (QA) on a D-Wave Quantum Annealer (D-Wave QA). This 
approach performs iterative QA measurements to optimize the eigenstates $\vert \psi \rangle$ without the 
derivation of a classical computer. The computational cost is $\eta M L$ for full eigenvalues $E$ and $
\vert \psi \rangle$ of the Hamiltonian $\hat{H}$ of size $L \times L$, where $M$ and $\eta$ are the 
number of QA measurements required to reach the converged $\vert \psi \rangle$ and the total annealing 
time of many QA shots, respectively. Unlike the exact diagonalized (ED) algorithm with $L^3$ iterations 
on a classical computer, the computation cost is not significantly affected by $L$ and $M$ because 
$\eta$ represents a very short time within $10^{-2}$ seconds on the D-Wave QA. We selected the 
tight-binding $\hat{H}$ that contains the exact $E$ values of all energy states in two systems with 
metallic and insulating phases. We confirmed that the proposed QE algorithm provides exact solutions 
within the errors of $5 \times 10^{-3}$. The QE algorithm will not only show computational supremacy 
over the ED approach on a classical computer but will also be widely used for various applications such 
as material and drug design.
\end{abstract}

\pacs{71.10.Fd,71.27.+a,71.30.+h}
% 71.10.Fd    Lattice fermion models (Hubbard model, etc.)
% 71.27.+a    Strongly correlated electron systems; heavy fermions
% 71.30.+h    Metal-insulator transitions and other electronic transitions
\keywords{}
\maketitle

\emph{Introduction}---Fast computation to find the full eigenvalues $E$ and eigenstates $\vert \psi 
\rangle$ of a Hermitian 
matrix $\hat{H}$ of size $L \times L$ is a highly desirable technique which has been widely applied 
across various applications, including many-body electronic systems, electronic structures, novel 
material design and drug design. The full values of $E$ and $\vert \psi \rangle$ can be computed using 
an exact diagonalization (ED) method, where the computational cost demands iterations of $L^3$ on a 
classic computer based on bits~\cite{Weise2008}. Another approach using a classic computer is to employ 
an optimization 
approach. The lowest eigenvalue $E^{\text{Lowest}}$ and its $\vert \psi \rangle$ can be determined using 
the optimization equation $E^{\text{Lowest}} = \text{min}_{\vert \psi \rangle} \langle \psi \vert 
\hat{H} \vert \psi \rangle$, where $\vert \psi \rangle$ retains continuous variables. A gradient descent 
(GD) method is then employed to solve the continuous-variable optimization problem. The computational 
cost for determining $E^{\text{Lowest}}$ and its $\vert \psi \rangle$ requires the derivation of 
$N \times L$, where $N$ is the number of iterations required to reach 
the converged $E^{\text{Lowest}}$. $N$ generally increases with $L$. The excited 
eigenvalue and eigenstate are also computed using the same optimization method with the constraint of 
the orthonormal condition between the eigenstates. The overall computational cost for determining the 
full $E$ and $\vert \psi \rangle$ in all the energy states is approximately $N^2 \times L^2$. Therefore, 
it is difficult to calculate the full $E$ and $\vert \psi \rangle$ of $\hat{H}$ with massive $L$ in a 
feasible time using both ED and GD approaches on classic computer.

The variational quantum eigensolver (VQE) algorithm on gate-type quantum computers with qubit circuits 
is another promising tool for determining $E^{\text{Lowest}}$ and its $\vert \psi \rangle$ of $\hat{H}
$~\cite{Preskill2018,Higgott2019,Bharti2022}. These are measured by the operation of qubits on a circuit 
in combination with derivations on a classical computer. Many degrees of freedom for expressing 
continuous variables $\vert \psi \rangle$ require a deep circuit, which is accompanied with large 
quantum errors. Therefore, another qubit is required to minimize the larger quantum errors. Present 
gate-type quantum computers are only composed of approximately hundreds of qubits, which is insufficient 
to reduce quantum errors for large $\hat{H}$, even though they show limitless possibilities with 
increasing qubits in the future~\cite{Preskill2018}.

The quantum annealing (QA) approach on adiabatic quantum computers is another alternative for fast 
computation of optimization problems~\cite{Kadowaki1998,Johnson2011,King2022}. The QA approach measures 
the possible $\vert \psi^{\text{Binary}} 
\rangle$ with only the optimized binary configurations through the quantum adiabatic process in the 
transverse-field Ising model, where $\vert \psi^{\text{Binary}} \rangle$ is the optimized binary spin 
configuration. This has already been performed on D-Wave quantum 
annealer (D-Wave QA) with 5000 qubits, which can compete with classic computers in terms of 
computational speed and accuracy for combinatorial optimization problems of moderate 
size~\cite{Amin2018,Inoue2021}. 
It can be also employed as quantum simulators for exploring phase transition and dynamic 
behaviors~\cite{King2021(1),Kairys2020,King2021,Park2022,Irie2021,Albash2018,Kumar2024,Teplukhin2020}. 
On the other hand, the 
method only works well on binary combinatorial optimization problems and is limited to 
continuous-variable optimization problems. Therefore, an extension of the approach to 
continuous-variable optimization problems that can be applied to more important applications is required 
in D-Wave QA.

In this Letter, we describe the development of a full quantum eigensolver (QE) that is based on the 
extension of optimized binary configurations determined using QA on the D-Wave QA. This method 
determines the possible $E^{\text{QE}}$ and $\vert \psi^{\text{QE}} \rangle$ values with continuous 
variables in all energy states through iterative QA measurements. The computational cost is $\eta M L$ 
for full $E^{\text{QE}}$ and $\vert \psi^{\text{QE}} \rangle$ of $\hat{H}$ of size $L \times L$, where 
$M$ and $\eta$ are the number of QA measurements required to reach the converged $\vert \psi \rangle$ 
and the total annealing time of many QA shots, respectively. This means that, because $\eta$ is very 
short, $L$ and the other parameters in $\hat{H}$ do not significantly affect the computational time. We 
selected a one-dimensional (1D) ionic non-interacting tight-binding $\hat{H}$ whose exact $E$ is known 
in the entire energy spectrum. We measured the possible $E^{\text{QE}}$ and $\vert \psi^{\text{QE}} 
\rangle$ in various cases of both metallic and insulating phases using our QE algorithm for D-Wave QA. 
We compared them with the exact $E$ and confirmed that it provides an exact $E$ within the errors of 
$5 \times 10^{-3}$. We believe that the proposed QE approach exhibits computational superiority over ED 
with $L^3$ iterations and GD with $N^2 L^2$ derivations on a classical computer.

\begin{figure}
\includegraphics[width=1.0\columnwidth]{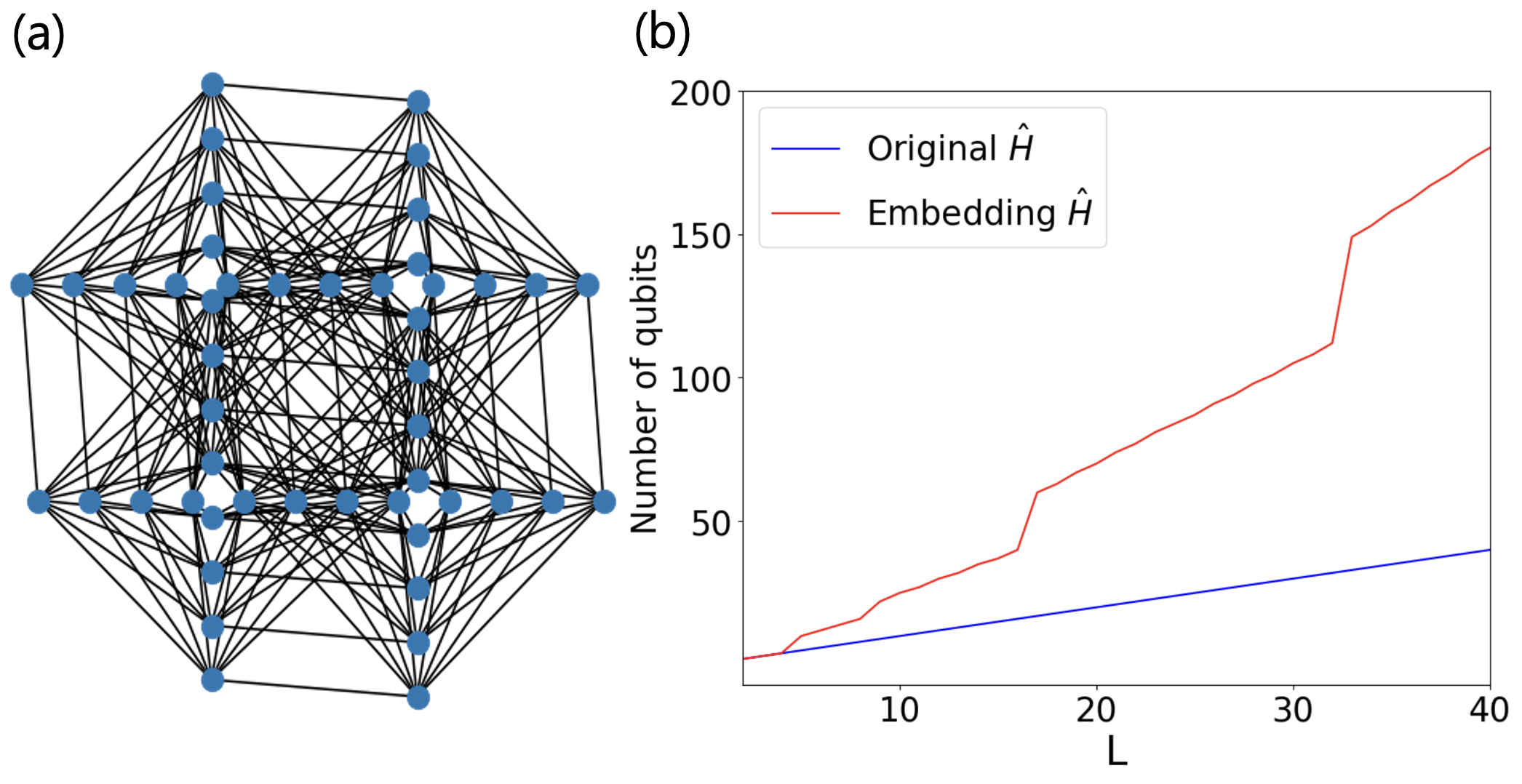}
\caption {\label{Fig1} (Color online) (a) Zephyr graph with $20$ couplers between qubits on
D-Wave QA Advantage2 prototype2.3. (b) Number of qubits as a function of $L$ for embedding 
Hamiltonian $\hat{H}$ required for quantum annealing measurement of the fully connected systems. The 
original $\hat{H}$ means the topology without the chains.}
\end{figure}

\emph{Algorithm of Quantum Eigensolver (QE)}---$E^{\text{Lowest}}$ and its $\vert \psi \rangle$ of 
$\hat{H}$ are determined by the optimization equation given as 
\begin{equation}
E^{\text{Lowest}} = \text{min}_{\vert \psi \rangle} \langle \psi \vert \hat{H} \vert \psi \rangle,
\label{Eq1}
\end{equation}
where $\vert \psi \rangle$ retains the continuous variable. The idea of our algorithm is to separate 
$\vert \psi \rangle$ into $\vert \psi \rangle = \vert \psi' \rangle + \vert \phi \rangle$.
Both $\vert \psi' \rangle$ and $\vert \phi \rangle$ are determined by iterative QA measurements. 
The detailed computational procedure is followed as: (i) We first determine $E'$ and 
$\vert \psi' \rangle$ in $E' = \text{min}_{\vert \psi' \rangle} \langle \psi' \vert \hat{H} \vert \psi' 
\rangle$ by QA measurement. The initial $\vert \psi' \rangle$ is only expressed as 
possible $\vert \psi^{\text{Binary}} \rangle$ that shows very rough solution of the optimized binary 
configuration. (ii) We modify $\hat{H}$ into $\hat{H}-E'I$, where $I$ is the identity matrix.
Now we rewrite $\vert \phi \rangle$ into $\vert \phi \rangle = \vert \psi \rangle - \vert 
\psi' \rangle$, and optimize $\vert \psi \rangle$ in $\hat{H}-E'\hat{I}$. The optimization equation 
for $\vert \psi \rangle$ is given as
\begin{equation}
E_{\vert \psi \rangle} = \text{min}_{\vert \psi \rangle} [(\langle \psi \vert - \langle \psi' \vert) 
\hat{H}^{(ii)} (\vert \psi \rangle - \vert \psi' \rangle)],
\label{Eq2}   
\end{equation}
where $\hat{H}^{(ii)}$ is $\hat{H}^{(ii)} = \hat{H}-E' \hat{I}$. $\vert \psi' \rangle$ were already 
known from QA measurement in (i) procedure. We measure $\vert \psi \rangle$ in Eq.~(\ref{Eq2}) through 
QA approach again. Note that $\langle \psi' \vert \hat{H} \vert \psi \rangle$ and $\langle \psi' \vert 
\hat{H} \vert \psi \rangle^\text{T}$ are only diagonal parts in the matrix of Eq.~(\ref{Eq2}). (iii) 
Next, $\vert \psi' \rangle$ is 
replaced into $\vert \psi \rangle$ measured in (ii) procedure, and (i) procedure 
is runing again. $\vert \psi' \rangle$ is no longer $\vert \psi^{\text{Binary}} \rangle$. (iv) This 
process is repeated until $\vert \psi' \rangle$ converges into $\vert \psi \rangle$ without assistance 
of derivation on classic computer.

\begin{figure}
\includegraphics[width=1.0\columnwidth]{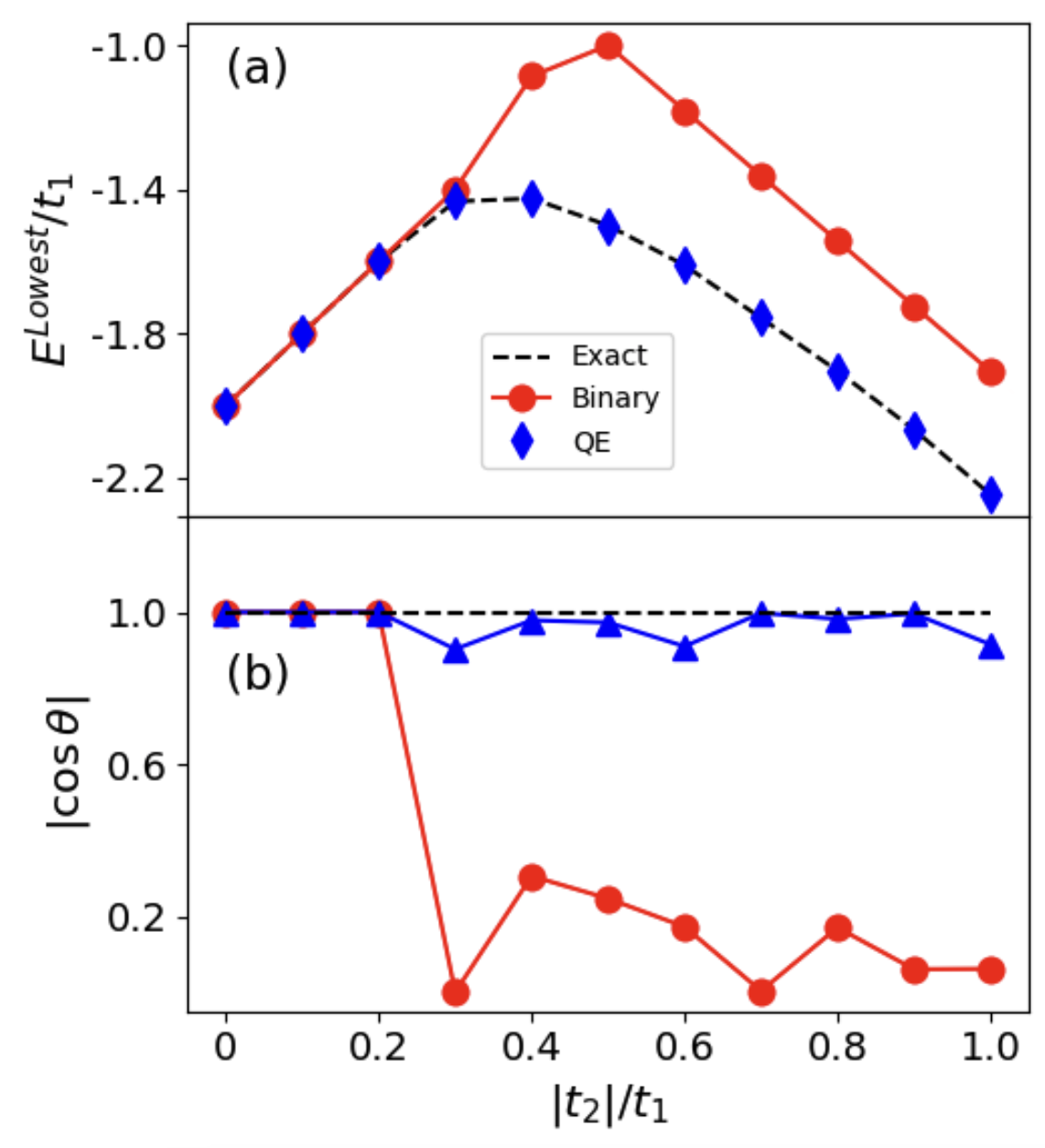}
\caption {\label{Fig2} (Color online) (a) The lowest eigenvalues $E^{\text{Lowest}}/t_1$ and (b) $\vert 
\cos \theta \vert$ as a function of $t_2/t_1$. The results are determined by the optimized binary 
configurations (Binary) and quantum eigensolver (QE) approaches. $\vert \cos \theta \vert$ is computed 
by $\vert \cos 
\theta \vert = \langle 
\psi^{\text{Exact}} \vert \psi^{\text{Compare}} \rangle$, where $\vert \psi^{\text{Compare}} \rangle$ is 
$\vert \psi^{\text{Exact}} \rangle$, $\vert \psi^{\text{Binary}} \rangle$ or $\vert \psi^{\text{QE}} 
\rangle$. Here, all eigenstates are normalized.}  
\end{figure}

The excited eigenvalue and its $\vert \psi \rangle$ are determined using the modified Hamiltonian 
$H^{\text{Excited}}$, which is given by $\hat{H}^{\text{Excited}} = \hat{H} - w \vert \Psi \rangle 
\langle\Psi \vert$, where $\vert \Psi \rangle$ is the eigenstate of a one-level-lower 
eigenvalue~\cite{Negre2022}. We 
impose an orthonormal condition on $\vert \psi \rangle$ and $\vert \Psi \rangle$. $w$ is a constant 
estimated as $w = h - l$, where $h$ and $l$ are the highest and lowest eigenvalues of $\hat{H}$, 
respectively. Finally, the excited $E$ and its $\vert \psi \rangle$ in $\hat{H}^{\text{Excited}}$ are 
computed using the same optimization method as described in Eq.~(\ref{Eq1}). The computational expense 
is $\eta M L$ for the full $E$ and $\vert \psi \rangle$ of $\hat{H}$. Notably, $L$ and other parameters 
in $\hat{H}$ do not significantly affect the computational time, contrary to the ED algorithm with $L^3$ 
iterations and GD optimization with $N^2 L^2$ algorithm on a classical computer, because $\eta$ is a 
very short time within $10^{-2}$ seconds for the D-Wave QA.

\emph{Hamiltonian for inspection of QE}---We selected the one-dimensional ionic $t_1-t_2$ tight-binding 
$\hat{H}$ that contains the exact $E$ in all energy states to demonstrate the accuracy and efficiency of 
our algorithm on the D-Wave QA. The Hamiltonian $\hat{H}$ is given by
\begin{align}
\hat{H}=&-t_1 \sum_{<i,j>} (c_{i}^{\dagger} c_{j} + \text{H.C}) - t_{2} \sum_{<i,j'>} (c_{i}^{\dagger} 
c_{j'} + \text{H.C}) \nonumber \\ &+ \sum_{i=0}^{L} \left[\Delta(-1)^{(i)} - \mu \right] n_{i},
\label{Hamiltonian}
\end{align}
where $c_{i}^{\dagger}$ and $c_{i}$ are the electron creation and annihilation operators at site $i$, 
respectively. H.C is a Hermitian conjugate. $t_1$ and $t_2$ are the nearest neighbors at site $j$ and 
next nearest-neighbor hopping at site $j'$, respectively. $L$ is the number of lattices under periodic 
boundary condition. $\mu$ and $\Delta$ are the chemical and ionic potentials, respectively.

\begin{figure}
\includegraphics[width=1.0\columnwidth]{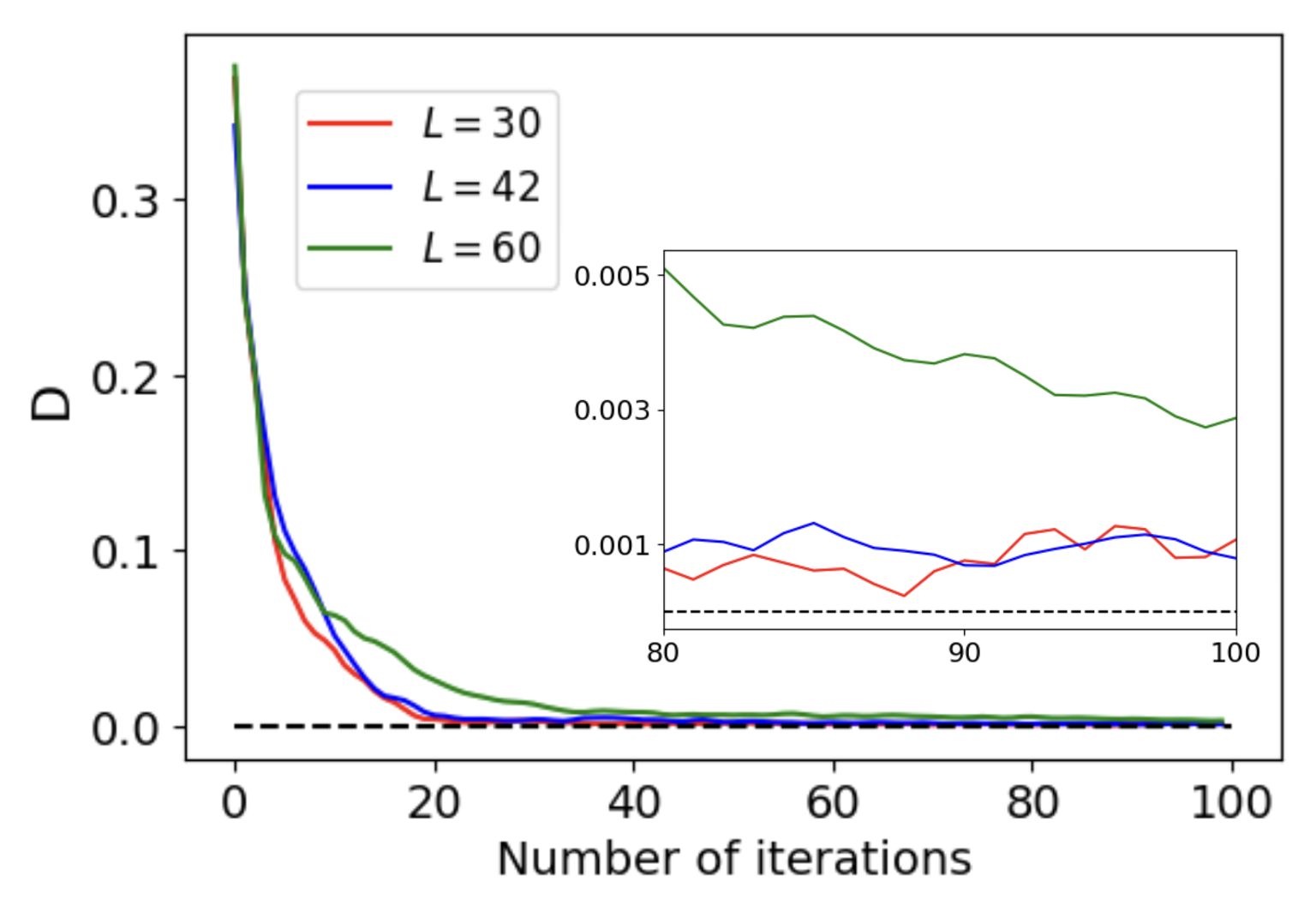}
\caption {\label{Fig3} (Color online) Deviations $D$ as a function of number of iteration for 
convergence in QE algorithm with different size $L$. $D$ are determined by $D=\vert E^{\text{Exact}}-
E^{\text{QE}} \vert$ at the lowest states with $t_2/t_1=1.0$. Inset shows the subtle $D$ after 
convergence of QA measurements using the QE approach. We confirmed that $D$ were below 
$5 \times 10^{-3}$ in all cases.}
\end{figure}

\emph{Quantum Annealing Measurements}---We used D-Wave QA Advantage2 prototype2.3, where the qubits are 
designed in a Zephyr structure, for the QA measurements, as shown in Fig.~\ref{Fig1} (a). Under Zephyr 
topology, each qubit is connected to $20$ different qubits via a coupler. The architecture of the 
original $\hat{H}$ (or $\hat{H}^{\text{Excited}}$) topologically matches that of a Zephyr graph by 
embeddings with chains of ferromagnetic (FM) order between qubits. Here, the embedding Hamiltonian 
$\hat{H}^{\text{Embedding}}$ is given by 
$\hat{H}^{\text{Embedding}} = \hat{H} + \hat{H}^{\text{Chain}}$, where $\hat{H}^{\text{Chain}}$ is the 
chain part required for embedding. More qubits are required than in the original 
ones to form chains. In addition, the elements in the excited $\hat{H}^{\text{Excited}}$ are completely 
occupied. We used the 'dwave.embedding.zephyr.find clique embedding' library provided by D-Wave Ocean 
Package for embedding of the fully connected systems. Fig.~\ref{Fig1} (b) shows the number of qubits as 
a 
function of $L$ for the original $\hat{H}$ and $\hat{H}^{\text{Embedding}}$. The optimal chain coupling 
for an appropriate FM order in the chains was determined using the method proposed by our group. The
annealing time was set to $10^{-4}$ seconds per a QA shot. The total annealing time $\eta$ with $100$ QA 
shots was set to $10^{-2}$ seconds.

\begin{figure}
\includegraphics[width=1.0\columnwidth]{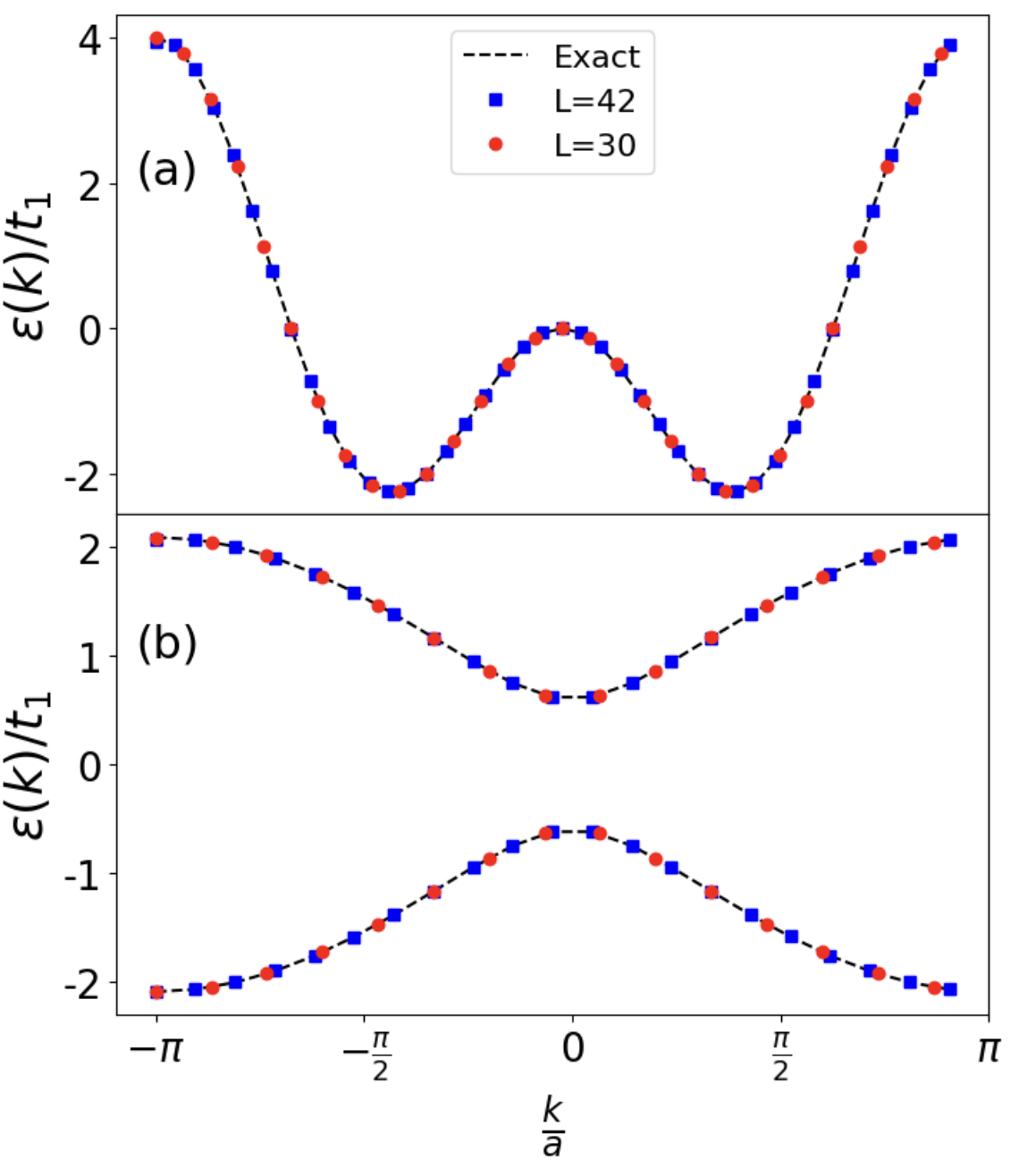}
\caption {\label{Fig4} (Color online) Energy dispersion $\epsilon(\bf{k})/t_1$ as a function of momentum 
$k/a$ for (a) metal with $\Delta/t_1=0.0$ and (b) insulator with $t_2/t_1=0.0$. Other parameters are 
$t_2/t_1=1.0$ and $\Delta/t_1=0.6$ for metal and insulator, respectively. The lattice constant $a$ is 
set to $1$.}
\end{figure}

\emph{The lowest eigenvalues and eigenstates on metallic phase}---We first measured $E^{\text{Lowest}}$ 
and its $\vert \psi \rangle$ of $\hat{H}$ with $\Delta=0.0$ as a function of $t_2/t_1$ using a QA with 
optimized binary configurations and QE approaches on the D-Wave QA. $E^{\text{Binary}}$ 
is determined using 
\begin{equation}
E^{\text{Binary}} = \text{min}_{\vert \psi^{\text{Binary}} \rangle} \langle \psi^{\text{Binary}} \vert 
\hat{H} \vert \psi^{\text{Binary}} \rangle.
\end{equation}
Fig.~\ref{Fig2} (a) shows $E^{\text{Lowest}}/t_1$ as a function of $t_2/t_1$. From $t_2/t_1=0.0$ to 
$t_2/t_1=0.3$, $E^{\text{Binary}}$ and $E^{\text{QE}}$ are equal to the exact $E^{\text{Exact}}$. After 
passing through this region, $E^{\text{QE}}$ remained equal to $E^{\text{Exact}}$ within errors of 
$5 \times 10^{-3}$, whereas $E^{\text{Binary}}$ was much higher than $E^{\text{Exact}}$. We also 
computed $\vert \cos \theta \vert$, which is given by the dot product of $\vert \psi^{\text{Binary}} 
\rangle$, $\vert \psi^{\text{QE}} \rangle$ and $\vert \psi^{\text{Exact}} \rangle$ to confirm the 
similarity of the eigenstates. Here, $\vert \psi^{\text{Exact}} \rangle$ was computed by the NumPy 
library. The results are present in Fig.~\ref{Fig2} (b). We confirmed that $\vert \cos \theta \vert$ 
between exact and QE eigenstates is nearly $1$. It means that $\vert \psi^{\text{QE}} \rangle$ and 
$\vert \psi^{\text{Exact}} \rangle$ agree well, as indicated by $E^{\text{Lowest}}/t_1$ in 
Fig.~\ref{Fig2} (a).

In Fig.~\ref{Fig3} we present the deviations $D$ as a function of the number of iterations in the QE 
algorithm for systems with $L=30$, $42$ and $60$ at $t_2/t_1 = 1.0$. Here, the deviations $D$ are 
computed by $D=\vert E^{\text{Exact}}-E^{\text{QE}} \vert$. Regardless of $L$, the convergence condition 
under which $D$ disappears was satisfied after $50$ iterations. It means that only $50 \times  
10^{-2}$ and $L \times 50 \times 10^{-2}$ seconds were enough to reach converged results in the QE 
algorithm of Eq.~(\ref{Eq1}) and to measure full $E^{\text{QE}}$ in all energy states, respectively. 
The inset of Fig.~\ref{Fig3} shows the subtle $D$ observed in the QA measurements using the QE 
approach, after convergence. We confirmed that $D$ were below $5 \times 10^{-3}$ in all cases.

\emph{Energy dispersion including information of the full eigenvalues of all spectrum}---Finally, we 
computed the energy dispersion $\epsilon ({\bf k})$ in two cases with metallic and insulating phases at 
$\mu=0.0$, because $\epsilon ({\bf k})$ includes $E$ in all energy states in the momentum space.
Here, $\epsilon ({\bf k})$ are expressed as $\epsilon ({\bf k}) 
= -2t_1 \cos {\bf k} - 2t_2 \cos 2{\bf k}$ and $\epsilon ({\bf k}) = \pm \sqrt{(-2t_1 \cos {\bf k})^2 + 
\Delta^2}$ for metal with $\Delta/t_1=0$ and insulater with $t_2/t_1=0$. We also employed $t_2/t_1=1.0$ 
and $\Delta/t_1=1.0$ for metallic and insulating phases, respectively. Here, $\bf k$ means the momentum 
space. The lattice constant $a$ is set to $1$. 
The results of $\epsilon(\bf k)$ that 
contain the full $E$ values for the entire spectra for the metallic and insulating phases of different 
$L$ values are shown in Figs.~\ref{Fig4}(a) and (b), respectively. We confirmed that the $E$ values 
determined using the QE method exactly matche exact $\epsilon ({\bf k})$ in both the metallic and 
insulating phases.

\emph{Conclusion}---The QA approach only provides $\vert \psi^{\text{Binary}} \rangle$ and its 
corresponding $E^{\text{Binary}}$. Thus, an extension of the optimized binary configuration is required 
to capture $\vert \psi \rangle$ with continuous variables. We developed a full QE based on 
iterative QA measurements on the D-Wave QA. The approach adjusted $\vert \psi \rangle$ correctly from 
the initial $\vert \psi^{\text{Binary}} \rangle$ without requiring derivation using a classical 
computer. The computational cost is $\eta M L$ for full $E$ and $\vert \psi \rangle$ for $\hat{H}$ with 
$L$. $\eta$ was set to be below the maximum value of $10^{-2}$ seconds. Thus, the 
computational time was not significantly affected by $L$ and $M$. This differs to the ED algorithm with 
$L^3$ iterations and the GD approach with derivations of $N^2 L^2$ on a classical computer. We 
considered two cases with metallic and insulating phases in the one-dimensional non-interacting ionic 
tight-binding $\hat{H}$ that contains the exact $E$ in the entire spectra to confirm the efficiency and 
accuracy of the QE algorithm on D-Wave QA. We determined that the iterations of the QA measurements 
using the QE method converged well. In addition, we confirmed that the QE method provided an exact $E$ 
within an error of $5 \times 10^{-3}$. Finally, we believe that the proposed QE will not only 
demonstrate computational supremacy over various numerical approaches on classic computers, but also be 
widely used for various applications, such as novel material and drug design.

\emph{Acknowledgements}---This work was supported by Institute of Information and communications 
Technology Planning Evaluation (IITP) grant funded by the Korean government (MSIT) (No. 
RS-2023-0022952422282052750001) and by the quantum computing technology development program of the
National Research Foundation of Korea (NRF) funded by the Korean government (Ministry of Science and
ICT(MSIT)) (No. 2020M3H3A1110365).

\end{document}